\documentclass[letterpaper]{article}
\usepackage{aaai2026}  
\usepackage{times}  
\usepackage{helvet}  
\usepackage{courier}  
\usepackage[hyphens]{url}  
\usepackage{graphicx} 
\usepackage{natbib}  
\usepackage{caption} 
\usepackage{booktabs}
\usepackage[table]{xcolor}
\usepackage{subcaption}
\usepackage{enumitem}
\usepackage{amsmath,amssymb}
\usepackage{placeins}
\usepackage{tikz}
\usetikzlibrary{arrows.meta,positioning,fit,calc,shapes.multipart,shapes.geometric,backgrounds}

\definecolor{softblue}{HTML}{E7F0FF}
\definecolor{softgreen}{HTML}{E9F8EF}
\definecolor{softorange}{HTML}{FFF3E0}
\definecolor{softgray}{HTML}{F5F7FA}
\definecolor{linegray}{HTML}{94A3B8}

\tikzset{
  box/.style={draw=black, rounded corners=2pt, very thick, fill=white, align=center, minimum height=8mm, inner xsep=6pt},
  pale/.style={draw=linegray, rounded corners=2pt, thick, fill=softgray, align=left, minimum height=7mm},
  tag/.style={draw=black, rounded corners=2pt, thick, fill=white, align=center, inner xsep=4pt, minimum height=5mm},
  arrow/.style={-Latex, very thick},
}

\nocopyright

\title{Human-in-the-Loop Interactive Report Generation for Chronic Disease Adherence}
\title{Human-in-the-Loop Interactive Report Generation for Chronic Disease Adherence}

\author{
    Xiaotian Zhang\textsuperscript{\rm 1},
    Jinhong Yu\textsuperscript{\rm 2},
    Pengwei Yan\textsuperscript{\rm 3},
    Le Jiang\textsuperscript{\rm 4},
    Xingyi Shen\textsuperscript{\rm 5},
    Mumo Cheng\textsuperscript{\rm 4},
    Xiaozhong Liu\textsuperscript{\rm 2}\thanks{Primary contact: xliu14@wpi.edu}
}
\affiliations{
    \textsuperscript{\rm 1}Northeastern University, United States\\
    \textsuperscript{\rm 2}Worcester Polytechnic Institute, United States\\
    \textsuperscript{\rm 3}Zhejiang University, China\\
    \textsuperscript{\rm 4}Shanghai Hongqiao Community Health Service Center, China\\
    \textsuperscript{\rm 5}Shanghai Zhujiajiao Town Community Health Service Center, China
}

\begin{document}
\maketitle
\vspace{-1.0em}

\begin{abstract}

Chronic disease management requires regular adherence feedback to prevent avoidable hospitalizations, yet clinicians lack time to produce personalized patient communications. Manual authoring preserves clinical accuracy but does not scale; AI generation scales but can undermine trust in patient-facing contexts. We present a clinician-in-the-loop interface that constrains AI to data organization and preserves physician oversight through recognition-based review. A single-page editor pairs AI-generated section drafts with time-aligned visualizations, enabling inline editing with visual evidence for each claim. This division of labor (AI organizes, clinician decides) targets both efficiency and accountability. In a pilot with three physicians reviewing 24 cases, AI successfully generated clinically personalized drafts matching physicians' manual authoring practice (overall mean 4.86/10 vs. 5.0/10 baseline), requiring minimal physician editing (mean 8.3\% content modification) with zero safety-critical issues, demonstrating effective automation of content generation. However, review time remained comparable to manual practice, revealing an \emph{accountability paradox}: in high-stakes clinical contexts, professional responsibility requires complete verification regardless of AI accuracy. We contribute three interaction patterns for clinical AI collaboration: bounded generation with recognition-based review via chart-text pairing, automated urgency flagging that analyzes vital trends and adherence patterns with fail-safe escalation for missed critical monitoring tasks, and progressive disclosure controls that reduce cognitive load while maintaining oversight. These patterns indicate that clinical AI efficiency requires not only accurate models, but also mechanisms for selective verification that preserve accountability.

\end{abstract}

\section{Introduction}
\begin{figure*}[t]
\centering
\includegraphics[width=\textwidth]{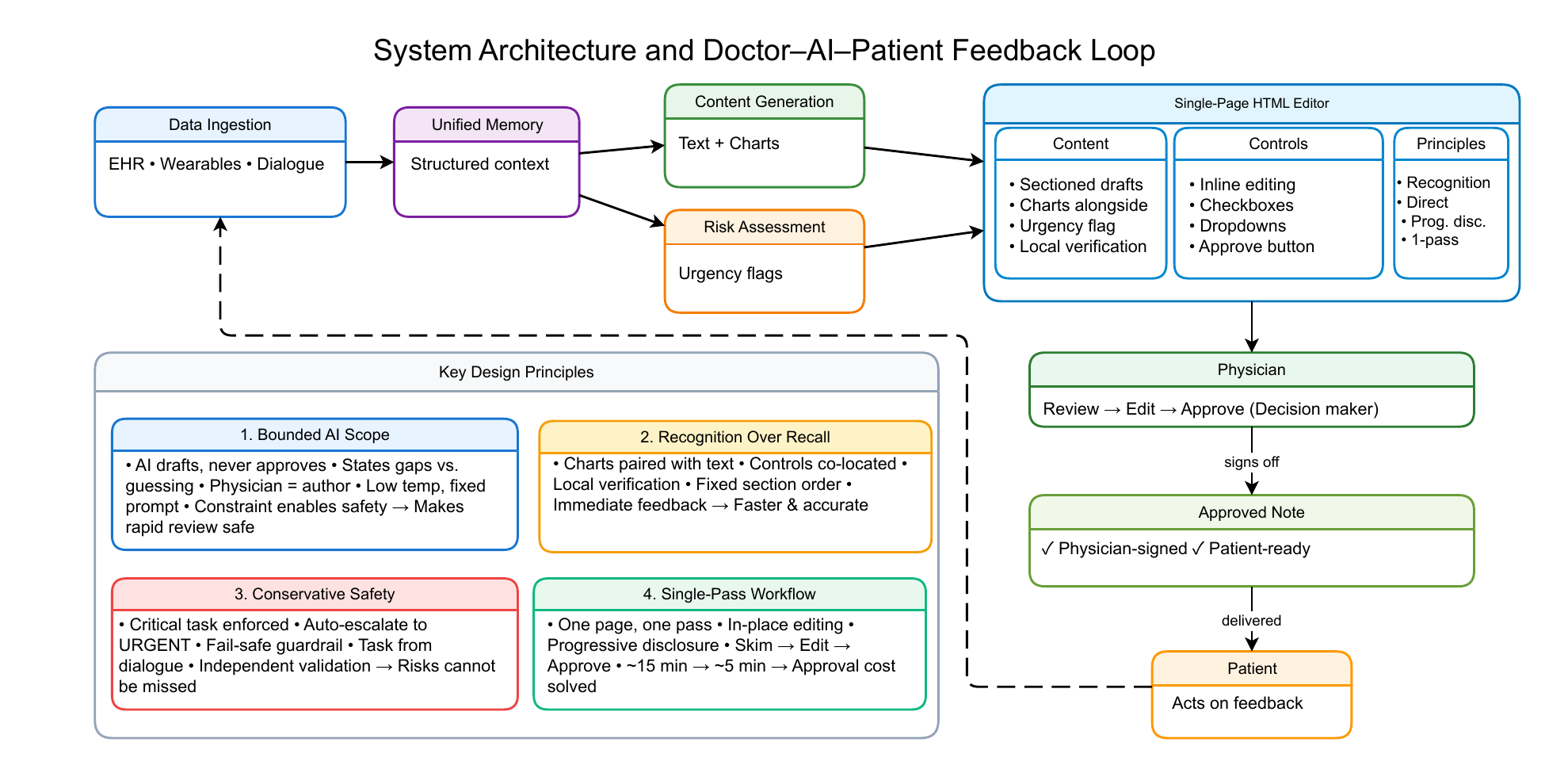}
\caption{\textbf{System architecture and feedback loop.} Data flows from patient through unified memory to parallel AI processing: content generation creates sectioned drafts, while risk assessment assigns urgency flags. The single-page HTML editor integrates both for physician review. Left panel shows four design principles. Right side shows the doctor-AI-patient feedback cycle.}
\label{fig:workflow}
\end{figure*}
Clinicians must turn fragmented records, device trends, and patient dialogue into actionable feedback, yet time pressure often yields generic notes that are hard to act on. Medication non-adherence remains widespread, driving preventable hospitalizations and costs~\cite{jimmy2011patient}. Documentation already consumes substantial clinical time~\cite{sinsky2016allocation}, leaving little capacity for personalized patient communication. Manual notes preserve nuance but do not scale; automated summaries scale but risk errors that undermine trust~\cite{lee2024prospects}. Existing human-in-the-loop tools frequently shift burden from drafting to review through multi-screen workflows, making \emph{approval}, rather than generation, the bottleneck.

We present an interaction design that reframes AI assistance as \emph{bounded preparation} rather than autonomous generation. The system deliberately divides labor: the AI organizes routine inputs (medications, device trends, dialogue) into structured drafts using a fixed template of \emph{what happened, why it matters, what to do next} and explicitly marks data gaps. Clinicians review on a \emph{single page} where each text section is paired with supporting visualizations, supporting recognition-based verification. Controls are co-located with affected content, edits preview immediately, and the workflow is designed for \emph{single-pass} approval.

To reduce cognitive triage across multiple patients, the system automatically labels cases as \emph{urgent}, \emph{attention}, or \emph{stable} by analyzing vital trends/deviations and adherence patterns inferred from dialogue. Critically, we enforce a conservative safety rule: if disease-specific \emph{critical monitoring tasks} are missed (e.g., daily blood pressure checks for hypertensive patients; glucose monitoring for diabetics), the case is \emph{automatically escalated to urgent} regardless of other indicators. This fail-safe ensures time-sensitive risks cannot be overlooked while keeping clinicians as authors of record.

We contribute three interaction patterns for efficient yet accountable clinical AI:
\begin{itemize}[leftmargin=*,itemsep=2pt]
    \item \textbf{Bounded AI preparation with recognition-based review.} AI is constrained to data organization (drafting, never approving). A single-page interface pairs sections with charts for local verification, supporting recognition-based review through recognition over recall.
    
    \item \textbf{Attention management via visual urgency flags.} Automated labels (\emph{urgent/attention/stable}) surface time-sensitive cases from vitals and adherence, reducing triage burden without dictating review order.
    
    \item \textbf{Conservative safety through fail-safe rules.} Missing disease-specific critical tasks triggers automatic urgent escalation, showing how explicit guardrails enable efficiency without sacrificing safety in high-stakes settings.
\end{itemize}

This approach aligns with human-centered principles for clinical AI~\cite{cai2019human} and recent governance frameworks~\cite{guidance2021ethics}, keeping clinicians the authors of record. Our pilot reveals an \emph{accountability paradox}, where professional responsibility requires complete verification despite high AI accuracy, demonstrating that clinical AI efficiency demands not just better models, but mechanisms for accountability-preserving selective verification.
\section{Related Work}

Clinical summarization using LLMs shows promise for reducing documentation burden, but challenges persist around factual accuracy and hallucination in patient-facing text~\cite{mannhardt2024impact, Hains2025Discharge,Small2025HCSummary,Bednarczyk2025Scoping}. While models can generate readable summaries, they often lack the clinical nuance required for personalized care plans and thus require human oversight. Human-in-the-loop systems in healthcare have explored various ways to balance automation with clinical judgment, from diagnostic support tools to documentation assistants~\cite{cai2019human}, yet many implementations introduce complex, multi-screen workflows that paradoxically increase cognitive load rather than reducing it. Evidence from EHR event logs and time-motion observations further shows that documentation-related work already occupies a substantial portion of primary care effort~\cite{arndt2017tethered,sinsky2016allocation}, underscoring that any AI assistance must integrate with minimal friction (fewer screens, clicks, and handoffs) to be viable in practice.

The medication adherence challenge is well-documented: non-adherence is widespread in chronic disease and drives substantial preventable costs~\cite{jimmy2011patient,WHO2003Adherence,Osterberg2005NEJM,Brown2011WHOcares}. Meta-analytic results indicate that clear, personalized clinician-patient communication is associated with improved treatment adherence compared to generic instructions~\cite{zolnierek2009physician, Zomahoun2017MIAdherence, Joosten2008SDM}. Recent surveys of LLM use in practice emphasize opportunities for summarization, rewriting, and dialog support while also highlighting the need for guardrails, error detection, and human-in-the-loop governance to ensure safety and trust~\cite{yang2024harnessing}. Building on these insights, prior clinical AI work has largely focused on improving generation quality to reduce documentation burden. However, less attention has been paid to how data organization and interface design, rather than generation quality alone, can enable efficient yet accountable review when professional liability cannot be delegated. We address this gap by constraining AI to data organization and optimizing the interface for verification.

\FloatBarrier
\section{Method and Approach}
\begin{figure}[t]
\centering
\includegraphics[width=\columnwidth]{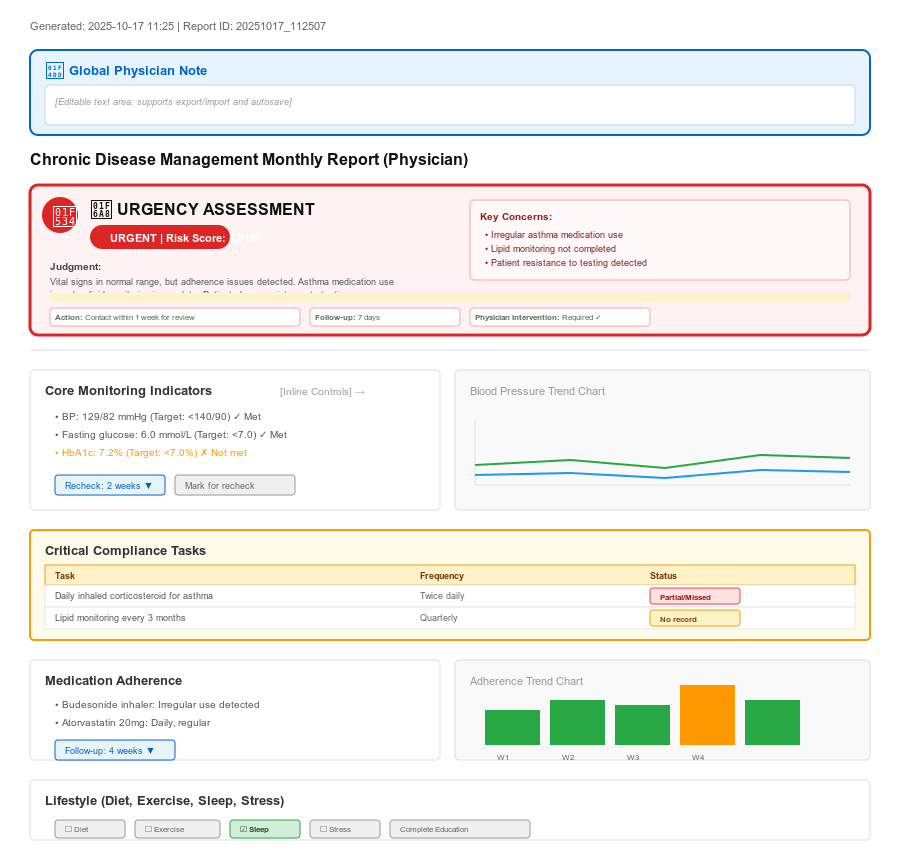}
\caption{\textbf{Physician interface for chronic disease management report.} }
\label{fig:ui}
\end{figure}
Our key insight is that deliberate AI constraints enable safety and efficiency in high-stakes domains. Unlike existing multi-stage review systems, we propose single-pass approval through careful task allocation and interface design. 

\textbf{Division of labor.} The model prepares; the physician decides; the patient receives an approved note. This doctor--AI--patient feedback loop establishes clear boundaries: the model does not approve or send anything to patients. The system architecture (Figure~\ref{fig:workflow}) implements this division through distinct components from data ingestion to physician-approved export.

\textbf{Inputs and draft.} For a reporting period, the system gathers routine items - medication lists and refills, basic device trends, and a brief dialogue - and prepares a short draft with a stable section order. The text appears as three small moves per topic: what happened, why it matters, and what to do next. Simple captions align with nearby plots. When data are missing, the draft states the gap rather than guessing.

\textbf{Technical details.}
Our system implements three key mechanisms. First, \textit{bounded AI preparation}: we constrain the LLM to draft structured text from routine inputs using a fixed template (what happened, why it matters, what to do), but the model never approves or sends content—physicians retain final control through single-pass review. We used Qwen3\textendash 8B with fixed decoding parameters (temperature = 0.7, max\_tokens = 1200) to ensure consistent and reproducible AI outputs across cases.

Second, \textit{automated urgency assessment}: to reduce cognitive triage across multiple patients, the system labels cases as urgent/attention/stable by analyzing vital trends, adherence gaps, and dialogue. These appear as color-coded badges at the top of each case, with urgent cases highlighted in red at the queue level, enabling physicians to quickly identify high-priority patients while maintaining discretion over review order. When LLM is unavailable, rule-based heuristics provide fallback classification. In practice, the LLM produces an initial urgency estimate, and a simple rule-based verifier applies secondary checks that may escalate or adjust the final label.

Third, \textit{conservative safety guardrails}: critically, if disease-specific monitoring tasks are missed (e.g., daily blood pressure checks for hypertension), cases automatically escalate to urgent regardless of other indicators. This prevents algorithmic optimism from hiding critical gaps that could lead to adverse events—a crucial safeguard when legal liability remains solely with physicians. The model generates draft text only and never produces autonomous clinical decisions; all final content is reviewed and approved by physicians.

\begin{figure}[t]
\centering
\resizebox{\columnwidth}{!}{%
\begin{tikzpicture}[node distance=6mm]
  \node[box, fill=softblue, minimum width=7.5cm] (a) {Load draft (sections + charts)};
  \node[box, fill=white, below=of a, minimum width=7.5cm] (b) {Read section summary};
  \node[box, fill=white, below=of b, minimum width=7.5cm] (c) {Check items (checkbox / dropdown / button)};
  \node[box, fill=white, below=of c, minimum width=7.5cm] (d) {Inline edit if needed};
  \node[box, fill=softgreen, below=of d, minimum width=7.5cm] (e) {Approve final note (physician-approved)};
  \draw[arrow] (a) -- (b);
  \draw[arrow] (b) -- (c);
  \draw[arrow] (c) -- (d);
  \draw[arrow] (d) -- (e);
\end{tikzpicture}}
\caption{\textbf{Structured review workflow.} One short pass: check essentials, make focused edits, approve and export.}
\label{fig:flow}
\end{figure}
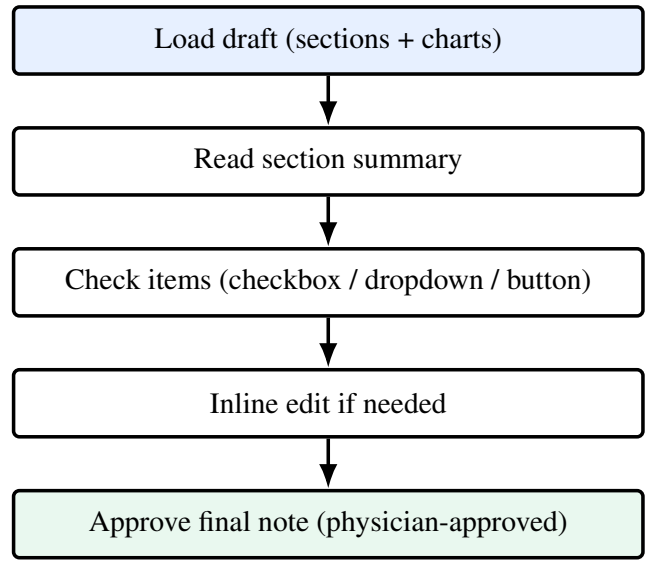

\textbf{Single-page review.} The physician works in one HTML page (Figure~\ref{fig:ui}). Sections are anchors for quick jumps. Sentences are edited where they appear. Small controls sit next to the items they affect, such as a checkbox to confirm medications, a short menu to set the follow-up interval, and a clear button to approve and export. Charts sit beside the sentences they explain, so checks are local and fast. A typical session follows the structured workflow in Figure~\ref{fig:flow}: open a case, skim from top to bottom, make focused edits, choose a follow-up interval, and approve.

\FloatBarrier

\textbf{Design rationale.} Summarizing routine inputs into a stable sectioned draft is a good fit for a model, while deciding what the patient should read remains a human task. The interface favors recognition over recall with a fixed section order and small, well-named choices; it supports direct manipulation by allowing in-place edits with immediate feedback; it applies progressive disclosure by keeping each block brief and inviting a short note only when needed. Scope remains tight: there are no extra trust badges, approval levels, or learning-from-edits features, and the implementation stays within the current editor and charting code.

\FloatBarrier
\section{Evaluation}

\subsection{Setup} 
We conducted a study with 3 physicians who collectively reviewed 24 generated chronic disease adherence reports. The cases were skewed towards higher-risk scenarios (14 urgent, 8 attention, and 2 stable), as these are the most critical for evaluating interventions where non-adherence poses the highest risk of preventable hospitalization.

After reviewing each case, physicians completed a structured questionnaire evaluating 12 quality dimensions on a 1–10 Likert scale. The baseline for comparison (score of 5) represented physicians' current manual authoring practice, which produces high-quality reports but typically requires 15+ minutes per report. This baseline was deliberately set at the midpoint to acknowledge that manual authoring already achieves good clinical quality. Scores above 5 indicated enhancement beyond this already effective practice, while scores below 5 indicated degradation. The 12 dimensions were organized into three domains:

\textbf{Core medical judgment (Q1-5):} urgency assessment accuracy, intervention recommendations, critical task identification, clinical appropriateness, and risk rationale quality.

\textbf{Data and factual accuracy (Q6-8):} data completeness identification, chart information value, and adherence description accuracy.

\textbf{Workflow integration (Q9-12):} readiness for consultation, time effort saved, information location efficiency, and overall satisfaction.

Physicians also recorded editing scope (unmodified, less than 10\%, 10-30\%, more than 30\%) and identified any safety concerns level.

\subsection{Results}

\textbf{AI-generated drafts achieved clinician-level quality.}
Across all 24 cases, physicians rated the AI-generated drafts at a quality level comparable to their manual authoring practice. The overall mean score across the 12 quality dimensions was 4.86/10 (SD = 0.52), where 5.0 represents physicians’ established baseline. Performance was consistent across dimensions: adherence accuracy was the strongest (Q8: 5.25), while intervention recommendations were relatively weaker (Q2: 4.42). Editing effort was minimal, with physicians modifying only an estimated 8.3\% of content on average (95\% CI: 3.3–13.3\%). Reviewer-level means varied (Physician 1: 5.31; Physician 2: 4.28; Physician 3: 5.00), reflecting differences in individual tolerance for AI-generated text rather than systematic model errors. One case raised a minor concern that was corrected during review, and no safety-critical issues were identified.

\textbf{Adequate quality did not translate into time savings.}
Despite satisfactory content quality, physicians did not report significant improvements in perceived time savings (Q10). The pooled mean across all cases was 4.79 (SD = 0.83), not significantly different from the baseline of 5.0 (t(23) = -1.23, p = 0.233). Individual reviewers showed heterogeneous patterns: Physician 1 reported exact parity with manual practice (mean = 5.00), Physician 2 rated below baseline (mean = 4.00, t(7) = -3.74, p = 0.007), and Physician 3 trended above baseline without significance (mean = 5.38, t(7) = 1.43, p = 0.197). These findings illustrate an \textbf{accountability paradox}: even when AI drafts are accurate, physicians must still perform complete verification due to non-delegable clinical responsibility, limiting the system’s ability to meaningfully reduce review time.

\textbf{Interface features supported verification-oriented workflow.}
Participants reported that the chart–text pairing effectively supported section-level verification, reflected in strong ratings for information-location efficiency (Q11:5.08, SD = 0.65). Urgency flags assisted with triage by surfacing high-priority cases, and the single-page layout reduced context switching. However, overall review duration remained unchanged because full verification was still required. The reliability of the questionnaire was high (Cronbach's $\alpha$ = 0.89), indicating strong internal consistency in the 12 dimensions of the evaluation.

\begin{table}[t]
  \centering
  \small
  \begin{tabular}{lc}
    \toprule
    \textbf{Metric} & \textbf{Result} \\
    \midrule
    \multicolumn{2}{l}{\textit{Quality (1--10, baseline = 5)}} \\
    \quad Urgency assessment (Q1) & 5.04 \\
    \quad Intervention recommendations (Q2) & 4.42 \\
    \quad Critical task identification (Q3) & 4.62 \\
    \quad Clinical appropriateness (Q4) & 4.88 \\
    \quad Risk rationale (Q5) & 4.83 \\
    \quad Data completeness (Q6) & 4.75 \\
    \quad Chart information value (Q7) & 4.96 \\
    \quad Adherence accuracy (Q8) & 5.25 \\
    \quad Readiness for consultation (Q9) & 4.83 \\
    \quad Time effort saved (Q10; overall) & 4.79 \\
    \quad Information location efficiency (Q11) & 5.08 \\
    \quad Overall satisfaction (Q12) & 4.87 \\
    \midrule
    \multicolumn{2}{l}{\textit{Physician Variability}} \\
    \quad Q10 (Physician 1) & 5.00 \\
    \quad Q10 (Physician 2) & 4.00 \\
    \quad Q10 (Physician 3) & 5.38 \\
    \quad Overall mean (Q1--Q12) & 4.86 \\
    \midrule
    \multicolumn{2}{l}{\textit{Safety}} \\
    \quad Minor concerns & 1/24 \\
    \quad Safety-critical issues & 0/24 \\
    \quad Mean modification rate & 8.3\% \\
    \bottomrule
  \end{tabular}
  \caption{Pilot study results across 24 cases reviewed by 3 physicians (14 urgent, 8 attention, 2 stable). Q10 shows per-physician time-effort ratings.}
  \label{tab:eval}
\end{table}

\subsection{Discussion and Implications}

Our finding reveals a critical challenge for clinical AI: interaction design improvements cannot overcome accountability-driven verification requirements. Even with near-baseline quality scores (4.86/10) and only one minor concern across 24 cases, physicians felt compelled to conduct complete review because patient safety consequences preclude selective verification strategies. 

This accountability paradox extends beyond interface design. Current medical malpractice insurance often excludes AI-related incidents while covering human errors~\cite{thedoctors2024}, and physicians retain full liability when using AI tools~\cite{shumway2024medical}.  This gap reinforces mandatory complete verification regardless of AI accuracy.

Addressing this accountability paradox requires both technical mechanisms and systemic solutions. Future systems need features like confidence-based selective verification, graduated trust based on accuracy, and accountability-preserving shortcuts like section-level approval. Beyond technology, this also demands evolving legal and insurance frameworks that recognize AI-assisted practice. Finally, our urgency flagging results showed value for workflow optimization, suggesting AI's benefits may lie in attention management rather than per-task time reduction.

\section{Conclusion}
We presented a clinician-in-the-loop interface that bounds AI to data organization while enabling recognition-based review through single-page design, chart-text pairing, and automated urgency flags. Our pilot study with 3 physicians reviewing 24 cases demonstrated near-baseline clinical quality (mean 4.86/10) with minimal safety concerns, yet review time remained comparable to manual authoring.

This \textbf{accountability paradox}, where professional responsibility mandates complete verification regardless of AI accuracy, represents our key contribution. The finding challenges assumptions that better AI quality or interfaces will reduce physician workload in high-stakes domains. Instead, clinical AI efficiency requires explicit mechanisms for accountability-preserving selective verification.


\bibliography{aaai2026}

\end{document}